\documentclass[aps,prb,twocolumn,superscriptaddress,preprintnumbers]{revtex4}
\usepackage[colorlinks,bookmarks=false,citecolor=blue,linkcolor=red,urlcolor=blue]{hyperref}
\usepackage{epsfig}
\usepackage{tikz}
\usepackage{graphicx}
\usepackage{physics}

\usepackage{dcolumn}
\usepackage{bm}

\usepackage{amsmath,amssymb}

\makeatletter
\renewcommand*\env@matrix[1][*\c@MaxMatrixCols c]{%
  \hskip -\arraycolsep
  \let\@ifnextchar\new@ifnextchar
  \array{#1}}
\makeatother

\usepackage{appendix}

\begin{document}

\title{Skyrmion-like textures in superconductors with competing order}

\author{V. Saran}
\affiliation{Department of Physics, Indian Institute of Technology Madras, Chennai 600036, India}
\affiliation{The Institute of Mathematical Sciences, HBNI, C I T Campus, Chennai 600 113, India}
\author{Madhuparna Karmakar}
\affiliation{Department of Physics, Indian Institute of Technology Madras, Chennai 600036, India}
\author{R. Ganesh}
\affiliation{The Institute of Mathematical Sciences, HBNI, C I T Campus, Chennai 600 113, India}

\date{\today}

\begin{abstract}
Vortex cores in a superconductor can develop structure and manifest competing orders. In strong magnetic fields, the inter-vortex distance can become short enough for vortex cores to overlap, giving rise to long ranged textures. We show that spin orbit coupling (SOC) can modify the interaction between vortex cores to give rise to undulating competing order fields. We illustrate this using the $SO(3)$ theory of competing orders, wherein superconductivity and a scalar order form a three-component vector field. We consider a spherical surface with a radial magnetic flux that creates two vortices. We find stationary solutions where both vortex cores develop competing order in (a) the same sense, and (b) opposite senses. The latter represents a topologically stable vector configuration analogous to a single skyrmion. Its free energy is lowered when SOC is introduced, making it the ground state beyond a threshold SOC strength. 
We study this physics on the two dimensional plane using the attractive Hubbard model with Rashba SOC. Here, charge density wave (CDW) order competes with superconductivity. We find phases with long ranged, but spatially modulated, CDW order where the average CDW moment vanishes. In a wide range of parameters, CDW loses its rigidity as SOC lowers the energy cost for domain wall formation.
We discuss consequences for systems such as the cuprates where charge order develops without a sharp diffraction peak.   
\end{abstract}
\maketitle
\section{Introduction}

Superconductors with low-lying competing phases can develop structure within vortices, with competing phases revealing themselves in the core region. This phenomenon is now well established with several theoretical\cite{Arovas1997,Hu2002,Gao2011} and experimental\cite{vortexcoreAFM,Lake2001,Machida2015} studies. Ordering within cores can also have interesting macroscopic consequences. For instance, at high vortex densities, the core regions of adjacent vortices can overlap, leading to percolation of competing correlations. This can provide coherence and rigidity to the competing phase on top of the superconducting background, giving rise to `supersolidity'\cite{Wu2013,KarmakarPRB2017}. In this article, we demonstrate a different scenario where overlap between adjacent vortices leads to anti-correlated core order. This gives rise to interesting ordering textures with similarities to skyrmion crystallization in magnets. The ingredient that drives this physics is spin-orbit coupling which is known to be important in several families of superconducting materials\cite{Rau2016,Day2018,Smidman2017}. 

Phase competition and vortex core order have gained prominence due to recent experiments on charge order in underdoped cuprates\cite{Gabovich2010,Chang2012,Grissonnanche2014,Fradkin2015}. There is strong evidence for charge order from X-ray scattering\cite{Ghiringhelli2012}, STM\cite{Howald2003,Wise2008}, NMR\cite{Wu2013,Zhou2017}, thermal transport\cite{Grissonnanche2014} and sound velocity probes\cite{LeBoeuf2013}. However, there is no true long-ranged charge order. At strong fields, a transition from weak two-dimensional charge order to strong three-dimensional order has been seen\cite{Gerber2015,Jang2016,Chang2016}. However, both cases represent short-ranged order with the coherence length changing from $\sim10 $ to $\sim100 $ lattice spacings\cite{Gerber2015,Chang2016}. Despite several studies, we do not understand the mechanism that gives rise to local charge order but not long ranged coherence. Disorder has been suggested as a possible reason; however, strong evidence is still lacking\cite{Achkar2014}. Motivated by this intriguing question, we consider a somewhat exotic possibility in this article. We present a scenario where vortex-core-overlap leads to oscillatory textures where net charge order vanishes, perhaps with charge correlations losing their rigidity. 

Ordered textures have a long history in magnetism, with a resurgence of interest since the discovery of skyrmion crystal phases\cite{Nagaosa2013,Zhou2018,han2017skyrmions}. Several realizations are now known, typically attributed to a combination of spin orbit coupling, anisotropy and large magnetic fields. Textured phases have also been studied in quantum Hall ferromagnets\cite{Brey1996,Cote2010}, where interactions give rise to arrangements of `merons' and `antimerons'. As a meron-antimeron pair can be thought of as a skyrmion, these states can be seen as precursors of skyrmion crystals. In the $SO(3)$ theory of competing phases, a superconducting vortex may also be thought of as a meron\cite{KarmakarPRB2017}. 
In this language, the overlap-induced supersolid is a crystal of merons alone, without any anti-merons\cite{KarmakarPRB2017}. This provides the setting for the present study, where we demonstrate that SOC stabilizes crystals that resemble arrangements of merons and antimerons placed adjacent to one another.

\section{SO(3) theory of competing order}
\label{sec.SO3}

A powerful theoretical tool to study phase competition is to introduce extended vectors, with different order parameters for components. If the vector is constrained to have constant length, the suppression of one phase will automatically manifest competing phases. This approach was proposed for the cuprates with superconductivity (one complex order parameter) and antiferromagnetism (with three real components) forming a five-dimensional vector\cite{Demler2004}. This $SO(5)$ theory was shown to give rise to ordered vortex cores\cite{Arovas1997}. Recently, an analogous $SO(3)$ theory has been proposed by two of the current authors\cite{KarmakarPRB2017}. Here, superconductivity competes with a scalar order parameter, say charge order. 
A microscopic realization of the $SO(3)$ theory is found in  the attractive Hubbard model at strong coupling. In this case, the competing order is checkerboard charge density wave (CDW) order. The $SO(3)$ theory is described by the action 
\begin{eqnarray}
\nonumber \mathcal{L} = \frac{\chi}{2} \left| \left( \mathbf{\nabla} - \frac{2ie}{\hbar}\mathbf{A} \right) \Delta(\mathbf{r} ) \right|^2
+ \frac{1}{8\pi} \left( \mathbf{\nabla} \times \mathbf{A} \right)^2 \\
+ \frac{\chi}{2} \vert \mathbf{\nabla}{\rho}(\mathbf{r} ) \vert^2
-\vert \Delta (\mathbf{r} )\vert^2 - (1- \delta)\vert {\rho} (\mathbf{r} )\vert^2,
\label{eq.LG}
\end{eqnarray}
The order parameters $\Delta$ and ${\rho}$ represent superconductivity and the competing order (e.g., CDW) respectively. They are constrained to obey $\{\vert \Delta (\mathbf{r}) \vert^2 + \rho(\mathbf{r})^2 = c^2\}$, a constant length constraint. We will refer to $\rho$ as the CDW order parameter below, although it may correspond to any competing scalar ordering parameter. 

The parameter $\delta$ encodes asymmetry. It represents the energy difference between uniform configurations of the two orders, with $\delta>0$ corresponding to lower energy for the superconductor. In the Hubbard model at half-filling, this term is related to next-nearest neighbour hopping\cite{KarmakarPRB2017} ($\delta \sim t'^2$). When a magnetic field is introduced via the vector potential $\mathbf{A}$, the superconductor forms vortices with CDW order in the core regions. The vortex can be interpreted as a meron in terms of a pseudospin vector given by 
\begin{equation}
\vec{n} = \{\mathrm{Real}(\Delta),\mathrm{Imag}(\Delta),{\rho}\}.
\end{equation}
While going around the vortex at a distance, the pseudospins lie in the plane and wind by $2\pi$. At the vortex core, a non-zero $z$-component emerges to preserve the uniform length constraint. In place of a magnetic field, disorder can also be used to bring out competing order. A point-like impurity suffices to seed local CDW correlations over a superconducting background\cite{KarmakarJPSJ2017}.

\section{Two vortices on a sphere}
\label{sec.sphere}
We study $SO(3)$ theory on the surface of a sphere, a convenient high-symmetry geometry for the study of inter-vortex interactions. We assume a radial magnetic field induced by a magnetic monopole sitting at the centre. As shown by Dirac\cite{Dirac1931}, this leads to a total flux of 
$h/e$ 
through the sphere. As a superconducting vortex corresponds to a flux of $h/2e$, two vortices will be induced by this flux. The vortices will position themselves at antipodal points to minimize repulsion. We take the two vortices to be located at the north and south poles. For ease of calculation, we assume a strong type-II superconductor with a large value for the $\kappa$ coefficient. In this regime, we may take the magnetic field to be constant over the surface of the sphere.

\begin{figure}
\includegraphics[width=\columnwidth]{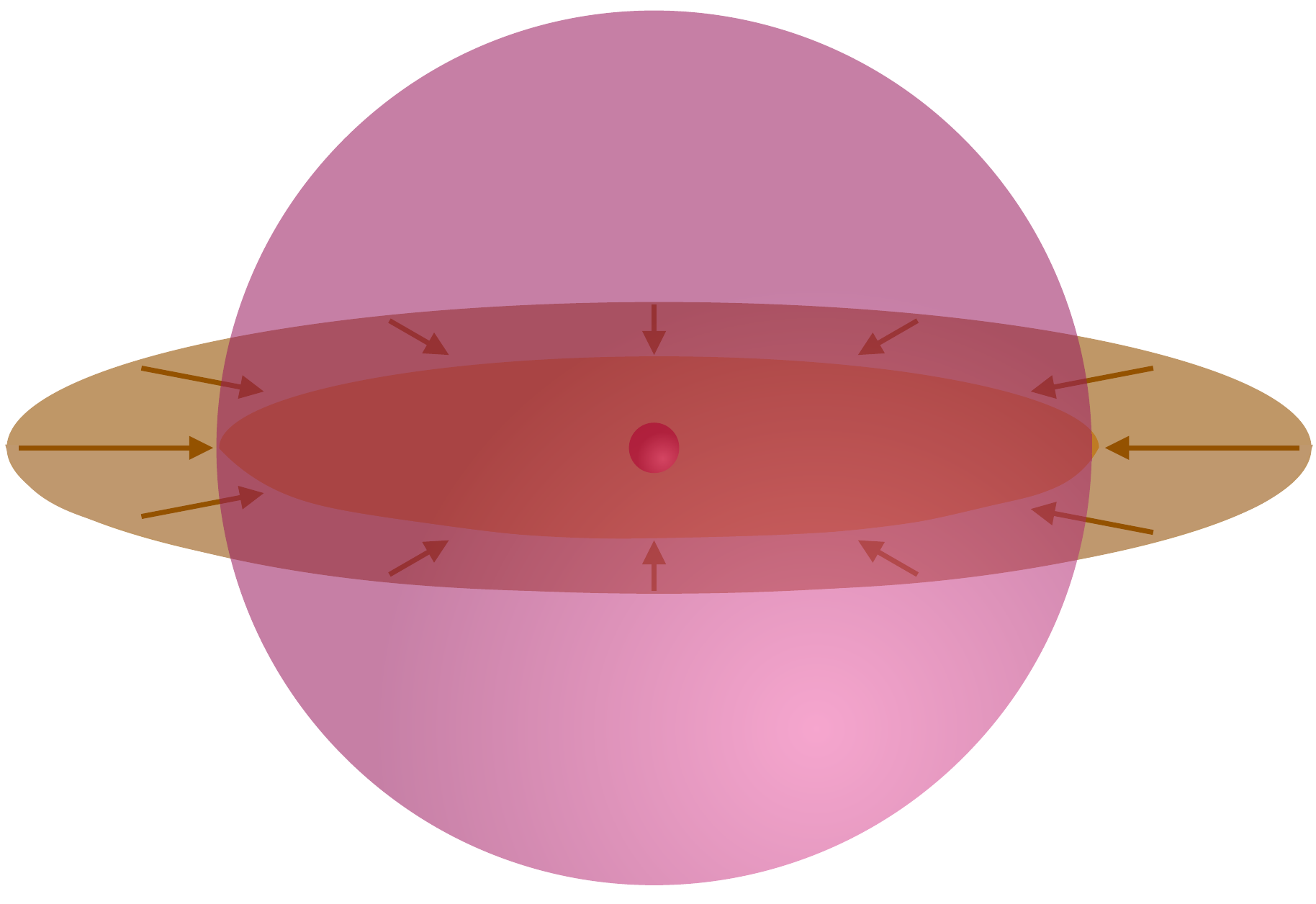}
\caption{Magnetic field of a monopole. We take the Dirac string to be flattened into a sheet in the equatorial plane.}
\label{fig.Dirac}
\end{figure}

The magnetic field is incorporated via a vector potential. As argued by Dirac\cite{Dirac1931}, it is impossible to define a vector potential so as to get a smooth non-zero magnetic flux. We will invariably find an anomalous flux corresponding to a `Dirac string' -- a solenoid that brings in flux into the sphere. This string is unphysical as it corresponds to a phase change of $2\pi$ for an electron on the surface. The position of the string can be varied by gauge transformations. Here, we make an unconventional gauge choice for reasons explained below, taking 
\begin{eqnarray}
\mathbf{A}(\theta,\phi) = \left\{ \begin{array}{c}
g  ~\hat{\phi} ~(1-\cos\theta)/{r\sin\theta}, ~~\theta <\pi/2 \\
-g ~ \hat{\phi}~ (1+\cos\theta)/r\sin\theta, ~~\theta >\pi/2
\end{array}\right. .
\label{eq.vecA}
\end{eqnarray}
Here, $\theta$ and $\phi$ represent coordinates on the sphere, viz., polar and azimuthal angles respectively. The radial coordinate is $r$, with the surface of the sphere assumed to be at $r=R$. The monopole charge, $g$, takes the smallest possible non-zero value, 
 $g= h/4\pi e$. 
By taking the curl of $\mathbf{A}(\theta,\phi)$, we find the magnetic field
\begin{eqnarray}
\mathbf{B}(r,\theta,\phi) = g \hat{r} /r^2 - \frac{2g}{r^2}\Big\{ \hat{r}  \sin\theta+ \hat{\theta}\cos\theta \Big\} \delta(r\cos\theta).
\end{eqnarray}
On the sphere's surface, we have a radial magnetic field of uniform strength in the northern ($\theta < \pi/2$) and southern ($\theta>\pi/2$) hemispheres. However, at the equator, we find an anomalous contribution with flux pointing inwards, towards the centre. This is shown in Fig.~\ref{fig.Dirac}. This corresponds to flattening the Dirac string into a Dirac sheet, with anomalous flux flowing inwards in all directions within the equatorial plane. The net anomalous flux is $h/e$, corresponding to a phase gain of $2\pi$ for an electron moving along a thin box enclosing the equator. 

We seek saddle point solutions of the action in Eq.~\ref{eq.LG} with the vector potential given in Eq.~\ref{eq.vecA}. We use the symmetry of the sphere to constrain our order parameters in the following manner. We take the superconducting order parameter to be
\begin{eqnarray}
\Delta(\theta,\phi) =  \left\{ \begin{array}{c}
f(\theta)~ e^{i\phi}, ~~\theta \leq \pi/2 \\
f(\theta)~ e^{-i\phi},~~ \pi/2 \leq \theta \leq \pi 
\end{array}\right. .
\label{eq.Deltadef}
\end{eqnarray}
We take the amplitude to be independent of $\phi$, reflecting azimuthal symmetry. However, the phase of the order parameter winds by $2\pi$ around each pole. This represents a two-vortex solution, with boundary conditions $f(0) = f(\pi) = 0$.
Around both poles, the winding corresponds to a flux pointing radially outward. 

Note that the superconducting order parameter is discontinuous at the equator. This is inevitable as the order parameter \textit{cannot} be smoothly defined on the surface of the sphere. This follows from Dirac's argument that the wavefunction of an electron cannot be smoothly defined on a closed surface that encloses a non-zero magnetic charge\cite{Dirac1931}. This holds for any charge-carrying scalar field that incurs an Aharanov-Bohm phase. In particular, it holds here for the charge-$2e$ superconducting pairing field on a sphere enclosing a monopole. This is an important aspect of our problem that is used to make an analogy with skyrmions in Sec.~\ref{ssec.skyrmion_analogy} below.
Here, the discontinuity can also be viewed as a consequence of having different gauges in the northern and southern hemispheres.

We take the ${\rho}$ order parameter to be given by
\begin{eqnarray}
\rho(\theta,\phi) = \pm\sqrt{c^2 - f^2(\theta)},
\label{eq.rho}
\end{eqnarray}
to account for the constant length constraint required by the $SO(3)$ theory. We assume azimuthal symmetry in this order parameter as well, with no dependence on $\phi$. There are two possible signs that can be assigned to $\rho$. We give sharper definitions for various cases below.

We have parametrized both order parameter fields, $\Delta$ and $\rho$, in terms of a single function, $f(\theta)$. We seek a function $f(\theta)$ that extremizes Eq.~\ref{eq.LG} and satisfies the boundary conditions, $f(0)=f(\pi)=0$. This is now precisely equivalent to the well-known problem in classical mechanics of determining a path from the principle of least action. 
A complication emerges from the vector potential in Eq.~\ref{eq.vecA} which has different functional forms in the northern and southern hemisphere. As a consequence, the Euler-Lagrange equation may take different forms in the two hemispheres. 
However, our gauge choice in Eq.~\ref{eq.vecA} provides a tremendous simplification, as it leads to \textit{identical} Euler-Lagrange equations in the two hemispheres. We may therefore treat this as a combined problem with boundary conditions at $\theta=0$ and $\theta=\pi$.

For gauge choices other than that of Eq.~\ref{eq.vecA}, we indeed obtain different Euler-Lagrange equations in the two hemispheres. The problem then has to be solved separately in each hemisphere, assuming a new boundary condition at the equator, $f(\theta=\pi/2) = f_0$.  
The solutions for different values of $f_0$ must be compared, and the one with the lowest energy chosen. With the gauge choice in Eq.~\ref{eq.vecA}, the value at the equator, $f(\pi/2)$, is automatically fixed by solving the problem on the entire sphere.

The Euler-Lagrange equation, in terms of a rescaled quantity $\tilde{f}(\theta)\equiv f(\theta)/c$,  is given by
\begin{eqnarray}
\nonumber \frac{\sin\theta \tilde{f}''}{1-\tilde{f}^2} &+& \frac{\tilde{f}\tilde{f}'^2 \sin\theta}{(1-\tilde{f}^2)^2} + \frac{\tilde{f}'\cos\theta}{1-\tilde{f}^2} \\
&+&\frac{\tilde{f}\sin\theta}{\xi^2} - \cos\theta\cot\theta \tilde{f} =0.
\label{eq.EL}
\end{eqnarray}
The details of the derivation are given in the Appendix~\ref{app.ELeqn}. Here, we have extracted a dimensionless scale, $\xi = \sqrt{\chi/2\delta R^2}$ from terms in the action. The action possesses an inherent length scale, $\sqrt{\chi/2\delta}$, which represents the size of the halo of competing order around a vortex\cite{KarmakarPRB2017}.
The quantity $\xi$ is the ratio of this length scale to the radius of the sphere.

We solve Eq.~\ref{eq.EL} using the shooting method, see App.~\ref{app.shooting}. 
We fix $f(0)=0$ and guess different values of $f'(0)$. This now becomes an initial value problem where $f(\theta)$ can be determined for all $\theta$ using the saddle point equation, using Runge-Kutta methods. We consider a solution to be physically acceptable if: (a) the boundary condition at $\theta=\pi$ is satisfied with $f(\pi)=0$, and (b) the amplitude of the pairing order parameter is symmetric about the equator, with $f(\theta)=f(\pi-\theta)$. Using this approach, we find three physical solutions for a given value of $\xi$. We discuss each case separately below.
 
\begin{figure*}
\includegraphics[width=0.48\columnwidth, height=1.6in]{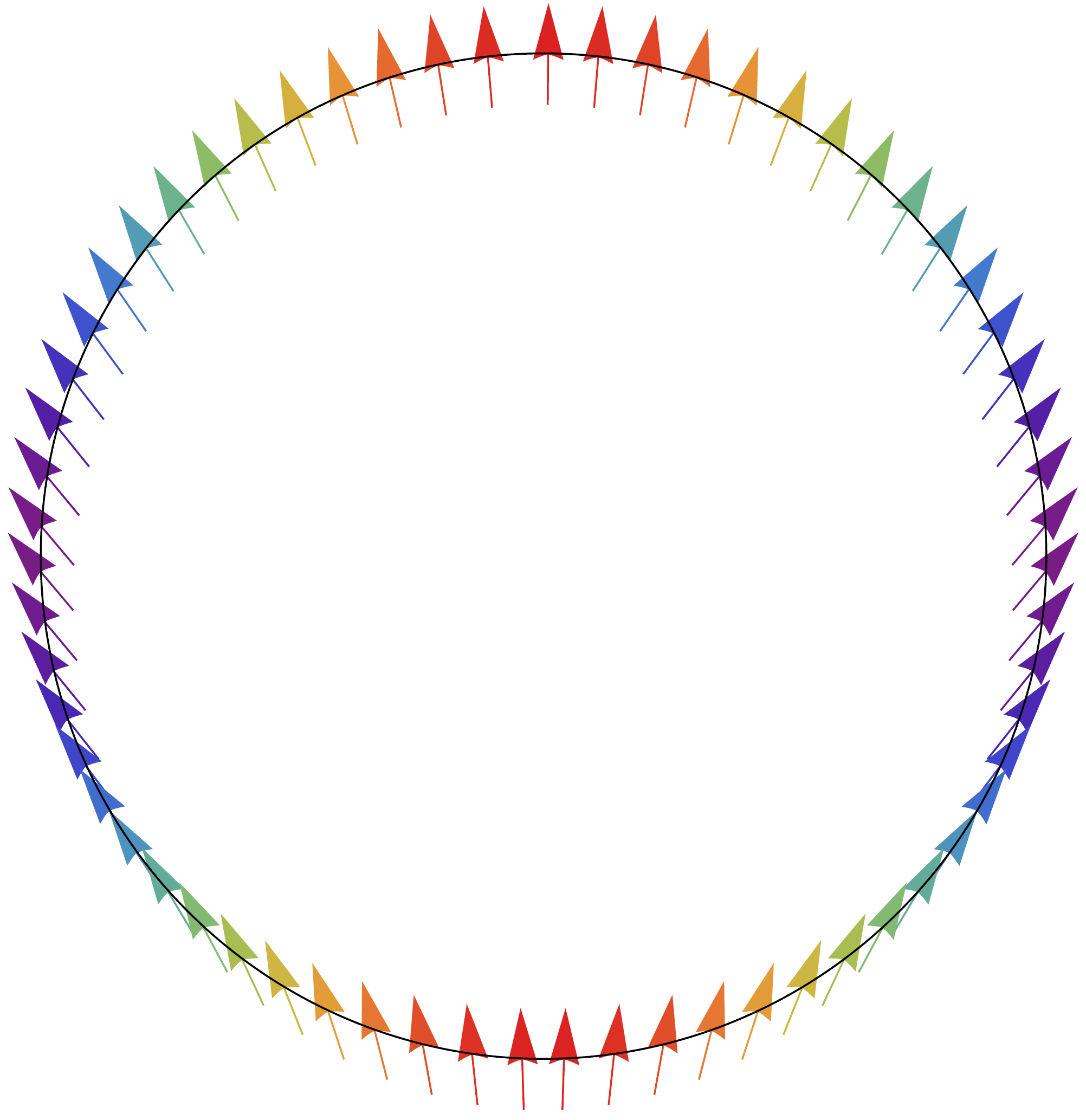}~~~~~
\includegraphics[width=0.7\columnwidth]{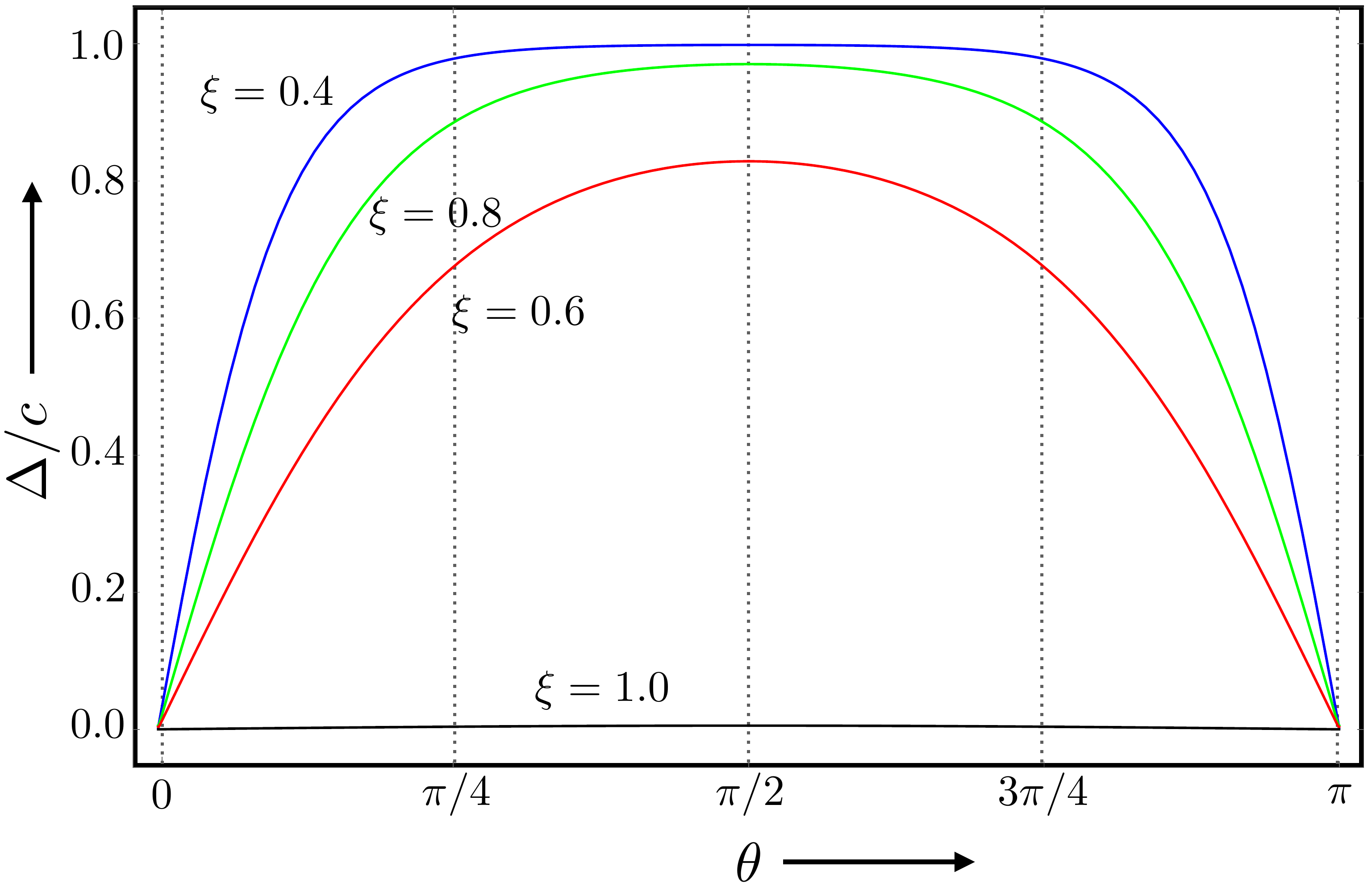}~~~~~
\includegraphics[width=0.7\columnwidth]{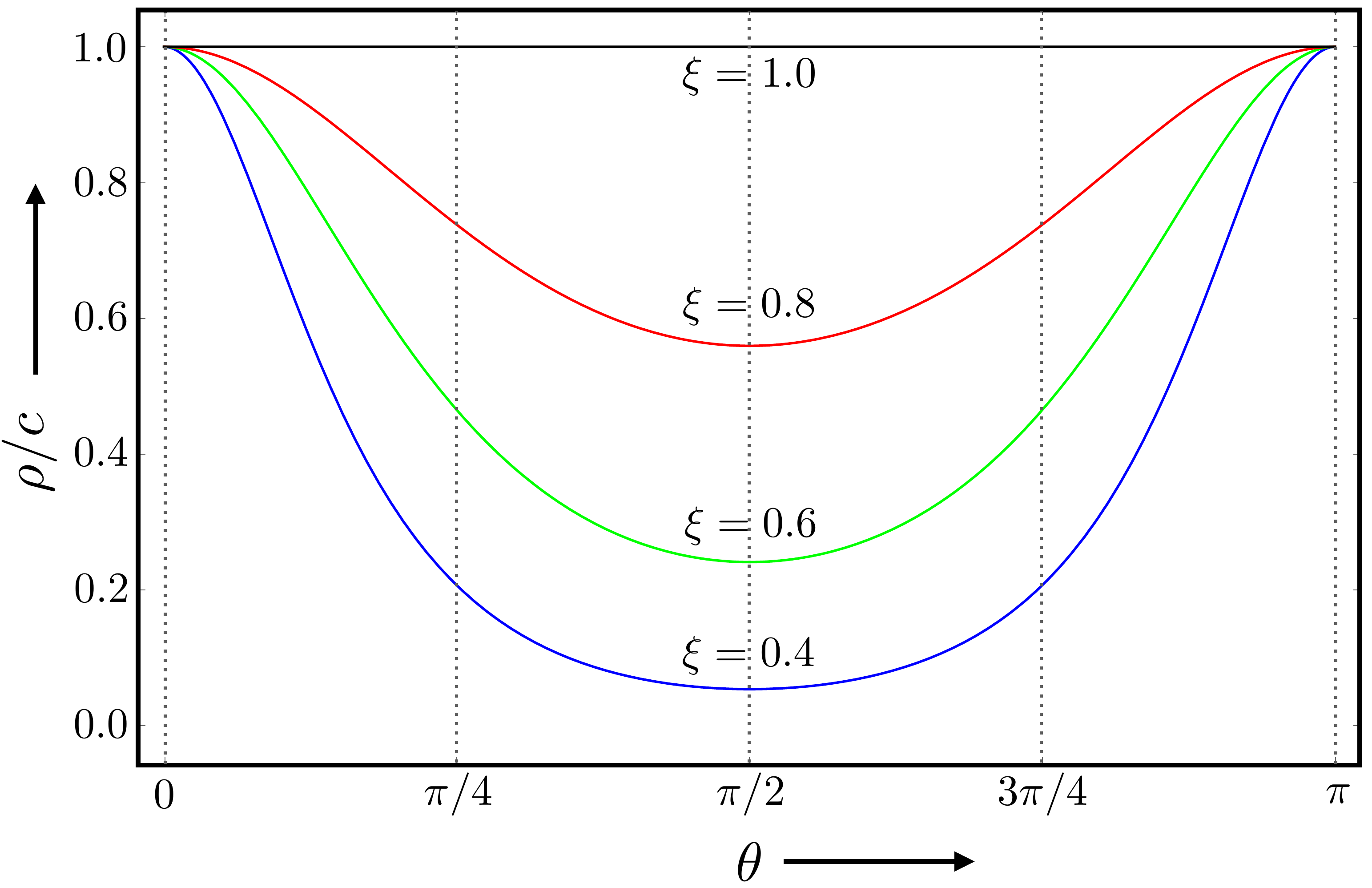}
\caption{Parallel solution. Left: A cross section of the sphere in the XZ plane showing the pseudospin orientation vs. $\theta$, for $\xi=0.9$.
The z-component of the pseudospin has the same sign everywhere, with strongest magnitude at the north and south poles. This corresponds to CDW order at both vortex cores developing in the same sense. Centre: Superconducting amplitude vs. $\theta$ for different $\xi$ values. Right: CDW order parameter vs. $\theta$ for different $\xi$ values. Note that CDW order is non-zero at the equator for $\xi \gtrsim 0.4$.
}
\label{fig.parallel}
\end{figure*}

\subsection{Saddle point solutions}

\underline{(a) Parallel solution}: In Eq.~\ref{eq.rho}, we have two possible signs for $\rho$. Here, we obtain the same sign (either $+$ or $-$) over the entire sphere in the solution. This is easiest to see in cases where the shooting method yields solutions with $f(\pi/2)<c$, i.e., the superconducting amplitude at the equator is less than the pseudospin length. In order to preserve the uniform length constraint, $\rho$ must be non-zero at the equator. For $\rho$ to be continuous and symmetric about the equator, it must necessarily have the same sign in both hemispheres.

The parallel solutions obtained for different values of $\xi$ are shown in Fig.~\ref{fig.parallel}. These solutions can be interpreted as follows: CDW order emerges at the two vortex cores centred at the poles. In both, CDW order has the same sign. For small $\xi$ values, the CDW halos around the vortices do not overlap. However, beyond a threshold value of $\xi$, they do. This leads to a non-zero value of $\rho$ at the equator. Consequently, the uniform length constraint forces the superconducting amplitude at the equator to be less than $c$.

\begin{figure*}
\includegraphics[width=0.48\columnwidth, height=1.6in]{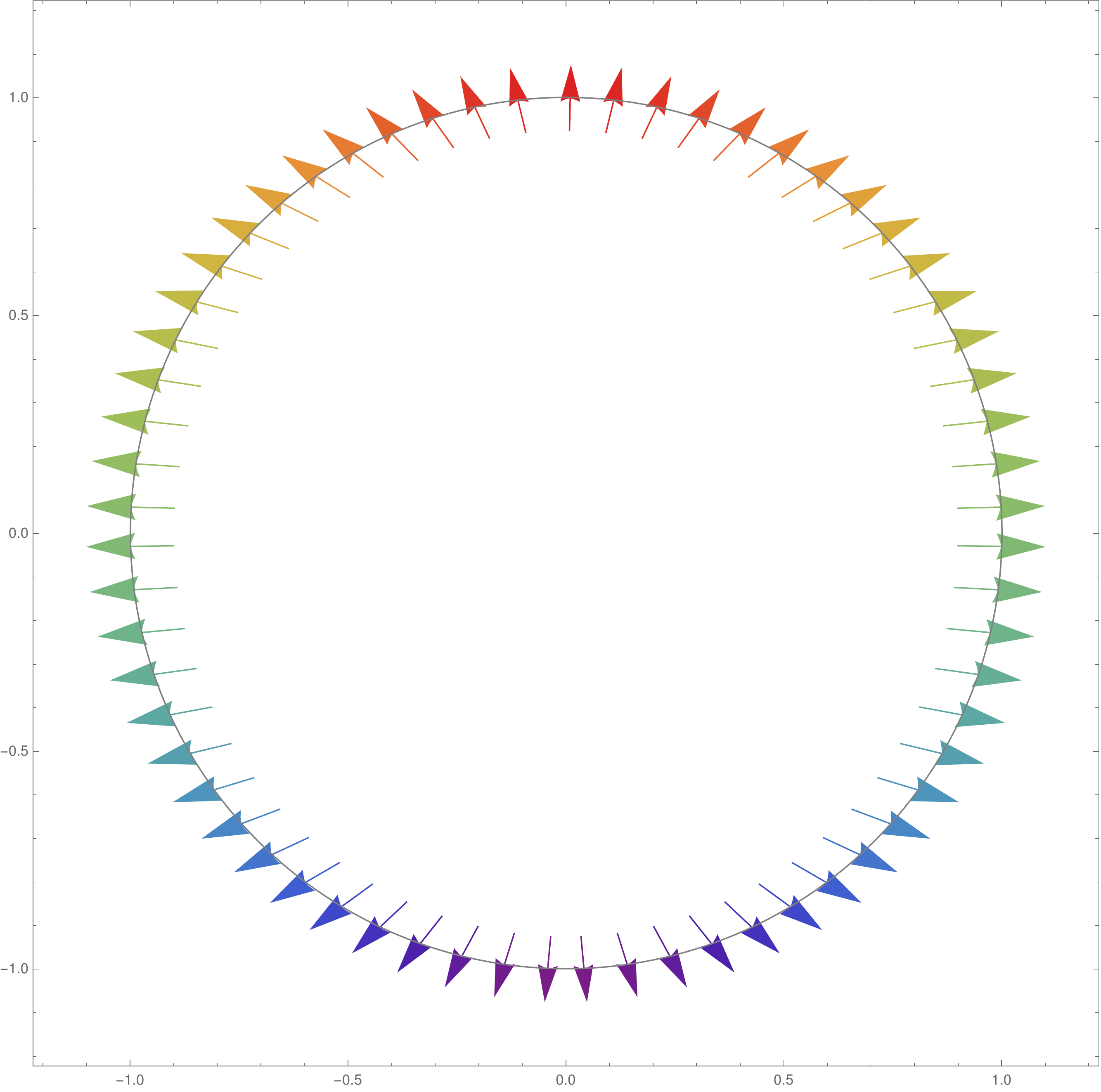}~~~~~
\includegraphics[width=0.7\columnwidth]{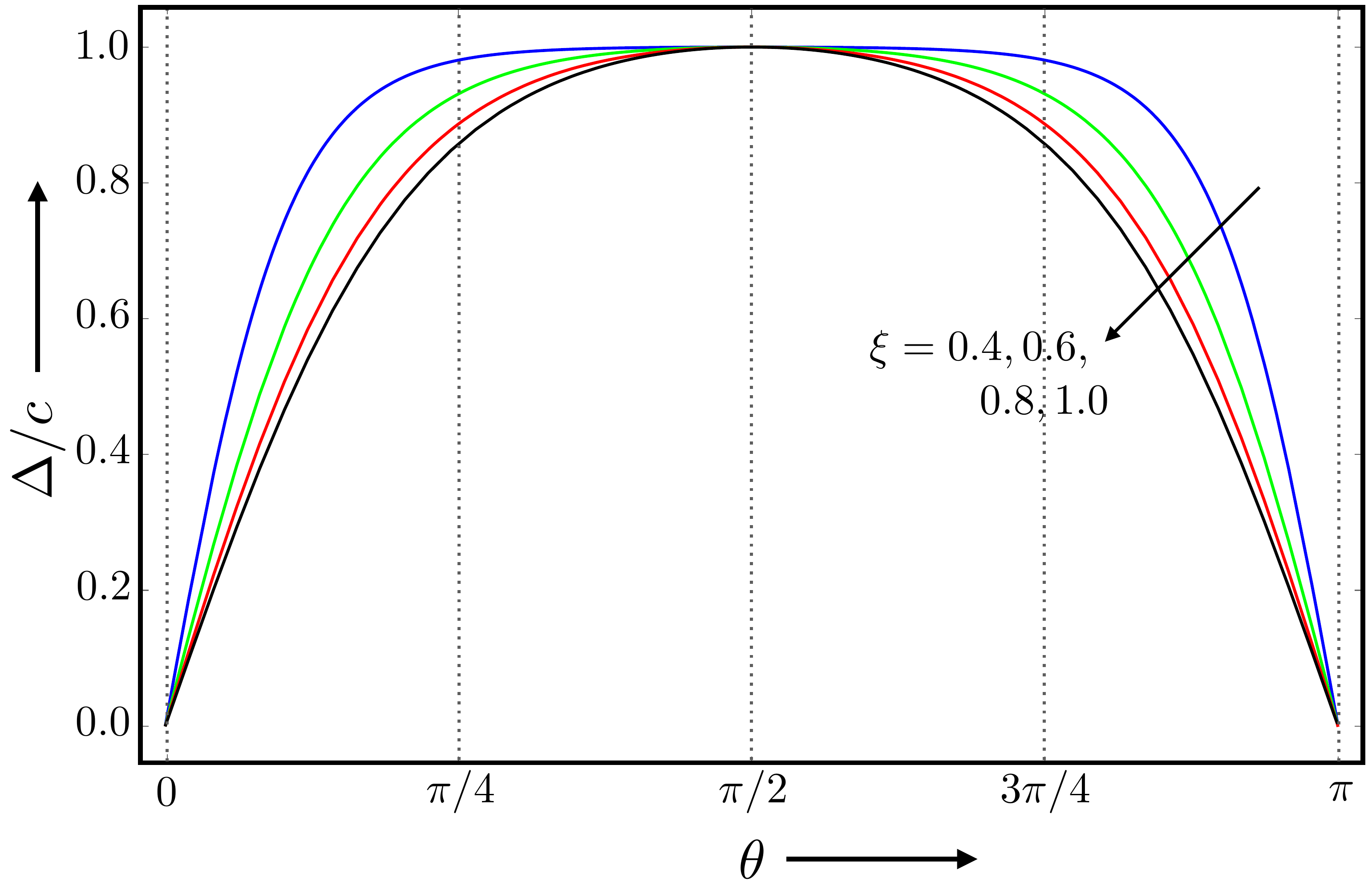}~~~~~
\includegraphics[width=0.7\columnwidth]{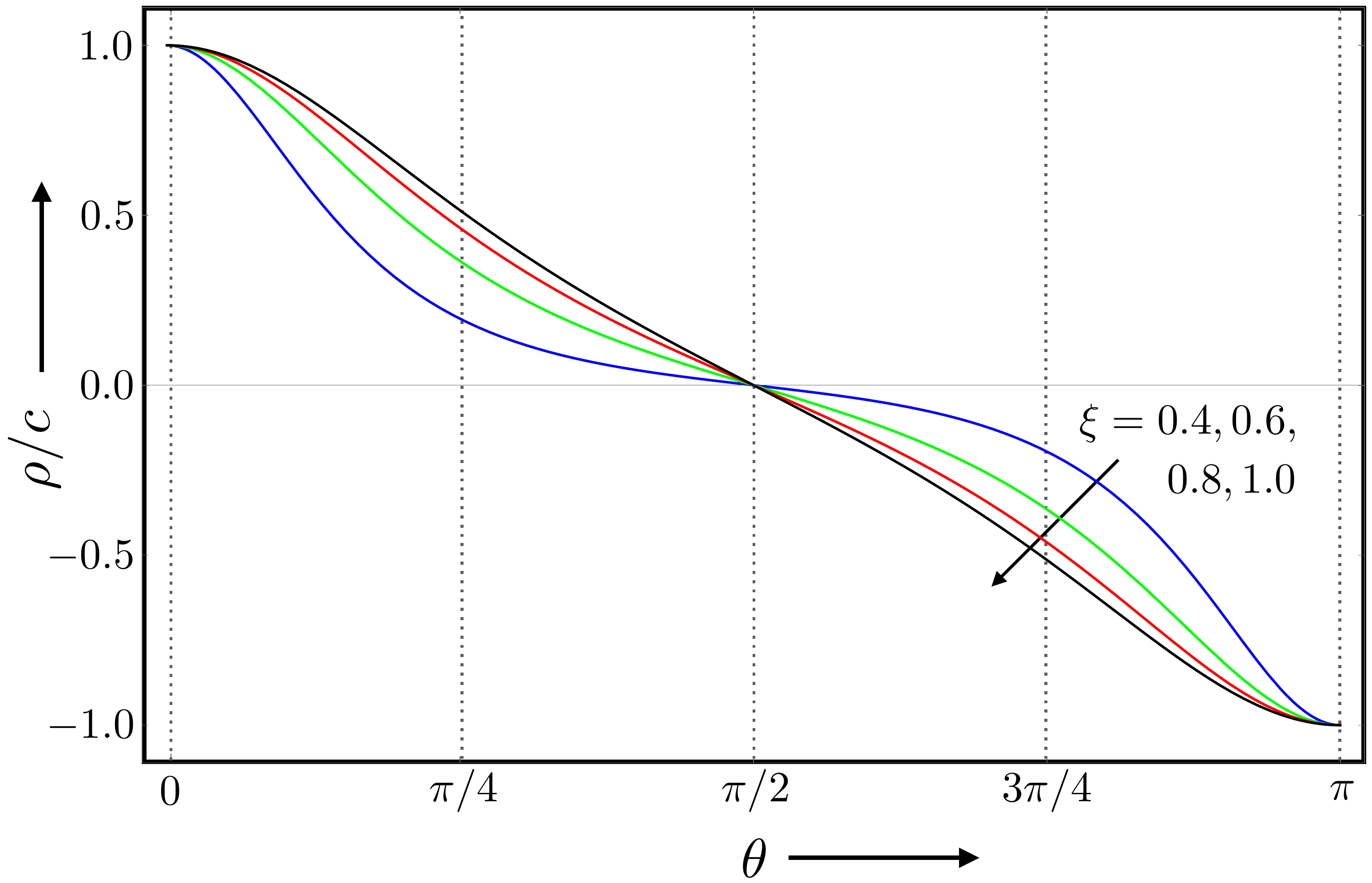}
\caption{Antiparallel solution. Left: A cross section of the sphere in the XZ plane showing the pseudospin orientation vs. $\theta$, for $\xi=0.9$. The z-component of the pseudospin has opposite signs in the two hemispheres, with highest amplitude at the poles. This corresponds to CDW order at the two vortex cores in opposite senses. Centre: Superconducting amplitude vs. $\theta$ for different $\xi$ values. Right: CDW order parameter vs. $\theta$ for different $\xi$ values. Note that CDW order is zero at the equator.}
\label{fig.antiparallel}
\end{figure*}

\underline{(b) Antiparallel solution}: In the form of $\rho$ given in Eq.~\ref{eq.rho}, we obtain the plus sign for the northern hemisphere and the minus sign for the southern hemisphere (or vice versa). We deduce this from the form of $f(\theta)$ obtained from the shooting method. We first note that this solution can only exist when $f(\theta)$ reaches the maximum possible value at the equator, with $f(\pi/2)=c$. 
As a result, $\rho$ vanishes at the equator to preserve the length constraint. This allows $\rho$ to be antisymmetric about the equator and yet continuous. In particular, this is the only possible solution when $f(\theta)$ decreases quadratically as we move away from the equator, i.e., $f(\pi/2 + \delta) \sim c - \alpha \delta^2$. To preserve continuity, $\rho(\theta)$ must go through zero linearly at the equator, with $\rho(\pi/2 + \delta)\sim \sqrt{2c\alpha}~ \delta$. This forces $\rho$ to take opposite signs in the two hemispheres.

The $\rho(\theta)$ profile is \textit{antisymmetric} under reflection about the equator, i.e., $\rho(\theta) = -\rho(\pi-\theta)$. 
Note that $f(\theta)$, the superconducting amplitude, is still symmetric with $f(\theta)=f(\pi-\theta)$. These solutions are shown in Fig.~\ref{fig.antiparallel}. They can be interpreted as follows. CDW emerges at both vortices, but with opposite sign. Continuity and symmetry force the CDW order parameter to vanish at the equator.

\underline{(c) CDW solution}: In addition to the above two solutions, there is always a trivial solution given by $f(\theta)=0$, where superconductivity is uniformly zero. To satisfy the uniform length constraint, we have uniform CDW order with maximum strength. There are two such solutions, corresonding to $\rho(\theta)=\pm c$. They are degenerate in energy. 

\subsection{Topological nature of the antiparallel solution}
\label{ssec.skyrmion_analogy}
We have discussed three saddle point solutions on the sphere above. The antiparallel solution is particularly interesting in the pseudospin language, as it is analogous to a `baby' skyrmion\cite{Piette1995}. This can be seen from Fig.~\ref{fig.antiparallel}(left) which shows the pseudospin profile on the XZ-plane cross section of the sphere. This shows a vector field that points radially outwards at every point. If we were to revolve this about the north-south axis, we would obtain a bonafide skyrmionic configuration on the sphere. However, there is a subtlety here.
We have shown the XZ plane in the figure, corresponding to azimuthal angle $\phi=0$ or $\pi$, as this allows for a continuous superconducting field \textit{without} a discontinuity at the equator. As seen from the definition of $\Delta(\theta,\phi)$ in Eq.~\ref{eq.Deltadef}, any other value of $\phi$ leads to a discontinuous jump at $\theta=\pi/2$.
Nevertheless, upon collecting the pseudospin values at all points on the sphere (as given by the antiparallel solution), we see that this covers a sphere in pseudospin space. In this sense, this solution is analogous to a skyrmion. 

The conventional definition of topological charge density (or skyrmion density or the Pontryagin index; see Ref.~\onlinecite{Nagaosa2013} for explicit definition) cannot be applied here as it only holds for smooth configurations of vector fields. Here, the in-plane components of the pseudospin (i.e., the real and imaginary parts of the superconducting order parameter) \textit{cannot} be defined smoothly on the sphere (see discussion below Eq.~\ref{eq.Deltadef}). 
Nevertheless, as with skyrmions, the antiparallel solution has topological stability. It has two vortices in the superconducting order parameter which cannot be removed by any smooth transformation. The CDW order parameter is tied to the vortex structure by the uniform length constraint of the $SO(3)$ theory. Notably, CDW order switches sign across the equator, where superconducting amplitude reaches its maximum value.
This line of CDW zeros cannot be removed by any smooth transformation that respects the uniform length constraint. For instance, it is impossible to smoothly deform the antiparallel solution into a parallel configuration. 

\subsection{Energy comparison}
\begin{figure}
\includegraphics[width=\columnwidth]{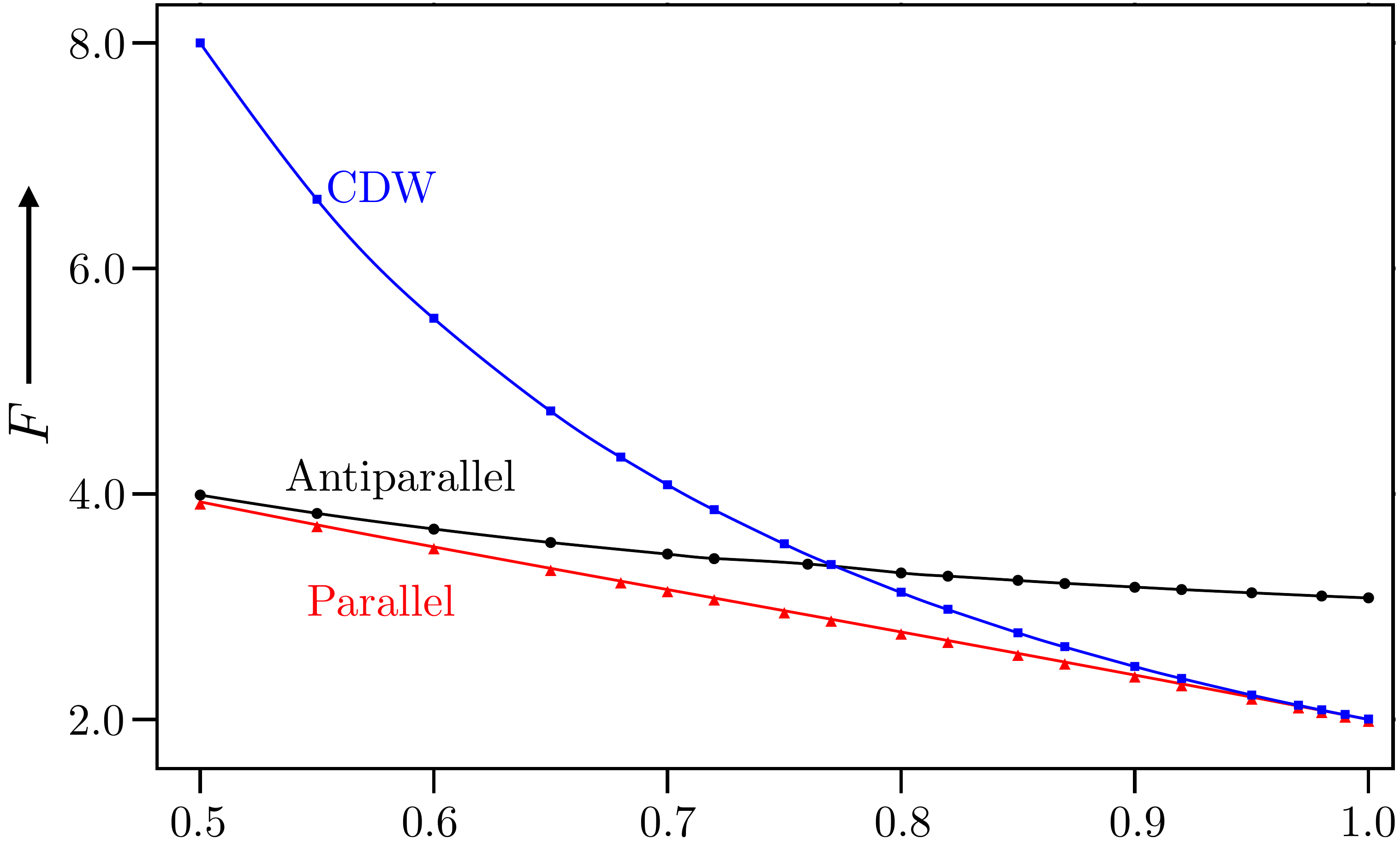}\\
\vspace{0.1in}
\includegraphics[width=\columnwidth]{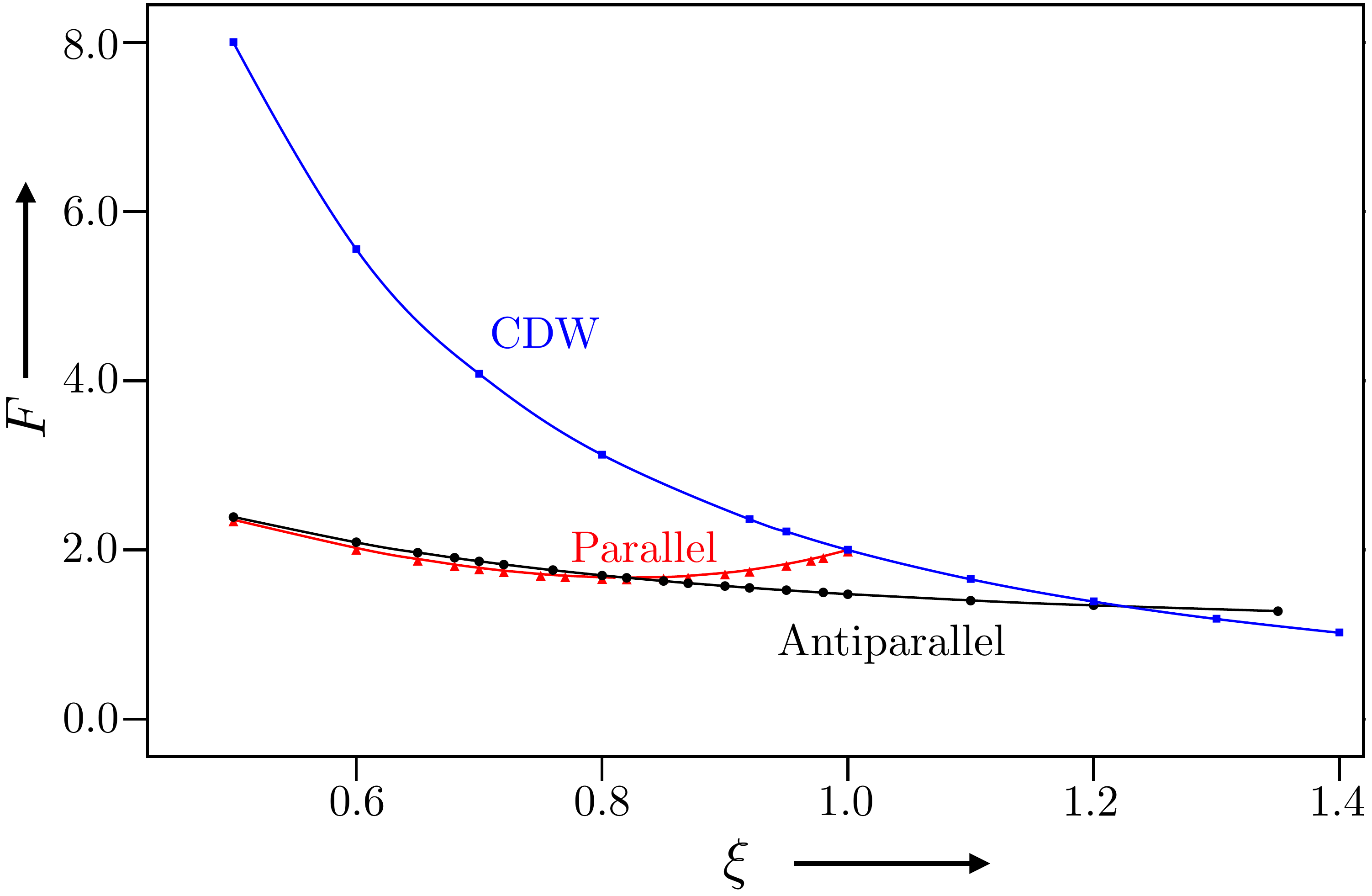}
\caption{Top: Free energies of the three saddle point solutions vs. $\xi$ within $SO(3)$ theory. Note that the parallel solution is always the ground state. Bottom: Free energies of the three solutions in the presence of SOC. The antiparallel solution becomes the ground state over a range of $\xi$ values. The data are for SOC strength given by $\lambda= -0.8$. 
}
\label{fig.energy}
\end{figure}

We have discussed three saddle point solutions that extremize the action. In Fig.~\ref{fig.energy} (top), we compare the free energies of these states vs. $\xi$. This corresponds to tuning $\delta$, the asymmetry parameter that controls the size of the CDW region in each vortex core, in Eq.~\ref{eq.LG}. As $\xi$ increases (or $\delta$ decreases), the CDW region shrinks. We find two threshold values of $\xi$ as seen in the figure: $\xi_{c1} \approx 0.35$ and $\xi_{c2}\approx 1$. For small $\xi$ (large $\delta$) values, with $\xi \lesssim \xi_{c1}$, the CDW regions around vortices are so small that they do not overlap. In this regime, the two CDW regions are independent. As a consequence, we find that the parallel and the antiparallel solutions have the same energies. For $\xi_{c1} \lesssim \xi \lesssim \xi_{c2}$, the two CDW regions overlap. As a result, the parallel solution is energetically preferable. This can be understood from the order parameter profiles shown in Figs.~\ref{fig.parallel} and \ref{fig.antiparallel}. The parallel solution suffers smaller gradients in both $\Delta$ and $\rho$ as compared to the antiparallel case. The gradient terms give a larger free energy contribution to the latter.  For $\xi \gtrsim \xi_{c2}$, the parallel solution and CDW solution become identical. If we track the parallel solution as we approach $\xi_{c2}$ from below, CDW order at each vortex grows progressively stronger. Simultaneously, superconductivity grows weaker so as to satisfy the uniform length constraint.  At $\xi_{c2}$, superconductivity vanishes leaving behind uniform CDW order.

\subsection{Role of spin orbit coupling}
\label{ssec.SOC}

We have argued that the anti-symmetric solution is skyrmion-like. However, we see from Fig.~\ref{fig.energy}(top) that it is not the ground state for any $\xi$ as it has higher free energy than the parallel solution. A suitable perturbation that reverses this energy hierarchy is given by
\begin{eqnarray}
L_{SOC} =
i \lambda  ~\hat{r} ~ \cdot \left\{
\mathbf{D} \Delta \times \mathbf{D} \Delta^*
\right\},
\label{eq.LRSOC}
\end{eqnarray}
where $\mathbf{D} =  \{ \mathbf{\nabla} -\frac{2ie}{\hbar }  \mathbf{A}\}$ is the gauge-invariant gradient operator. On a flat two-dimensional surface (the XY plane), the analogue of this term is $L_{SOC} =
i \lambda  ~\hat{z} \cdot \left\{
\mathbf{D} \Delta \times \mathbf{D} \Delta^*
\right\}$. Clearly, this term arises from symmetry breaking that differentiates the $z>0$ and $z<0$ regions, e.g., due to a perpendicular electric field. We designate this an SOC term as it may arise from Rashba spin-orbit coupling (RSOC). We adapt this to our spherical surface in Eq.~\ref{eq.LRSOC}.

The explicit form of this term is different in the northern and southern hemispheres. Remarkably, in both, this term only adds a total derivative to the action. Consequently, it does not modify the saddle point equation given in Eq.~\ref{eq.EL}. The three solutions discussed above continue to remain saddle point solutions. However, their free energies are modified by a boundary term (see details in Appendix~\ref{app.SOCsphere}). We find
\begin{eqnarray}
\delta F_{SOC} = 4\pi \lambda~ f^2(\pi/2),
\phantom{A}
\label{eq.FE_SOC}
\end{eqnarray}
wheres $f(\pi/2)$ is the amplitude of the superconducting order parameter at the equator, obtained in the corresponding solution. From Figs.~\ref{fig.parallel} and \ref{fig.antiparallel}, we see that the parallel and antiparallel solutions have different values of $f(\pi/2)$ when $\xi \gtrsim \xi_{c1}$. Physically, this corresponds to the regime where the two vortex cores overlap. In this regime, SOC changes the relative energies of these states. Remarkably, it retains the very same solutions while merely shifting their energies. We see that SOC with $\lambda<0$ lowers the energy of the antiparallel solutions. We note that SOC does not alter the free energy of the CDW solution as $ f_{CDW}(\pi/2)=0$.

The resulting behaviour of the free energy is shown in Fig.~\ref{fig.energy}(bottom) for $\lambda=-0.8$. SOC gives rise to a regime where the skyrmion-like antiparallel solution has the lowest free energy. This suggests that SOC could lead to skyrmionic textures in several families of superconductors which possess competing phases.

\section{Spin orbit coupling in the attractive Hubbard model}
In the discussion above, we have shown that SOC leads to a skyrmion-like configuration on the surface of a sphere. While the sphere is a convenient geometry for calculations, it does not capture the spatial structure of superconducting materials. For example, in the sphere considered above, we have two vortices that are neighbours of each other. Starting from one, we may approach the other by moving in any direction. This is very different from the spatial coordination in an Abrikosov vortex lattice on a flat two dimensional surface. 
Below, we study this more realistic geometry using the attractive Hubbard model on a square lattice, given by
\begin{eqnarray}
\nonumber H_{Hubb.} &=& -t \sum_{\langle ij \rangle,\sigma} c_{i,\sigma}^\dagger c_{j,\sigma} -t' \sum_{\langle\langle ij \rangle\rangle,\sigma} c_{i,\sigma}^\dagger c_{j,\sigma} +h.c. \\
&-&\mu \sum_{i,\sigma} c_{i,\sigma}^\dagger c_{i,\sigma} - U\sum_i c_{i,\uparrow}^\dagger c_{i,\downarrow}^\dagger c_{i,\downarrow} c_{i,\uparrow}.
\label{eq.Hubbard}
\end{eqnarray}
Here, $t$ and $t'$ represent nearest-neighbour and next-nearest-neighbour hopping on a square lattice. We have a chemical potential, $\mu$, and an on-site attractive interaction, $U$. For concreteness, we fix $t'=0.2t$ in all calculations presented here. 
As discussed in Sec.~\ref{sec.SO3}, this model, at half-filling and strong coupling ($U\gg t$), provides a microscopic realization of the $SO(3)$ theory of competing orders. 

\begin{figure}
\includegraphics[width=0.6\columnwidth]{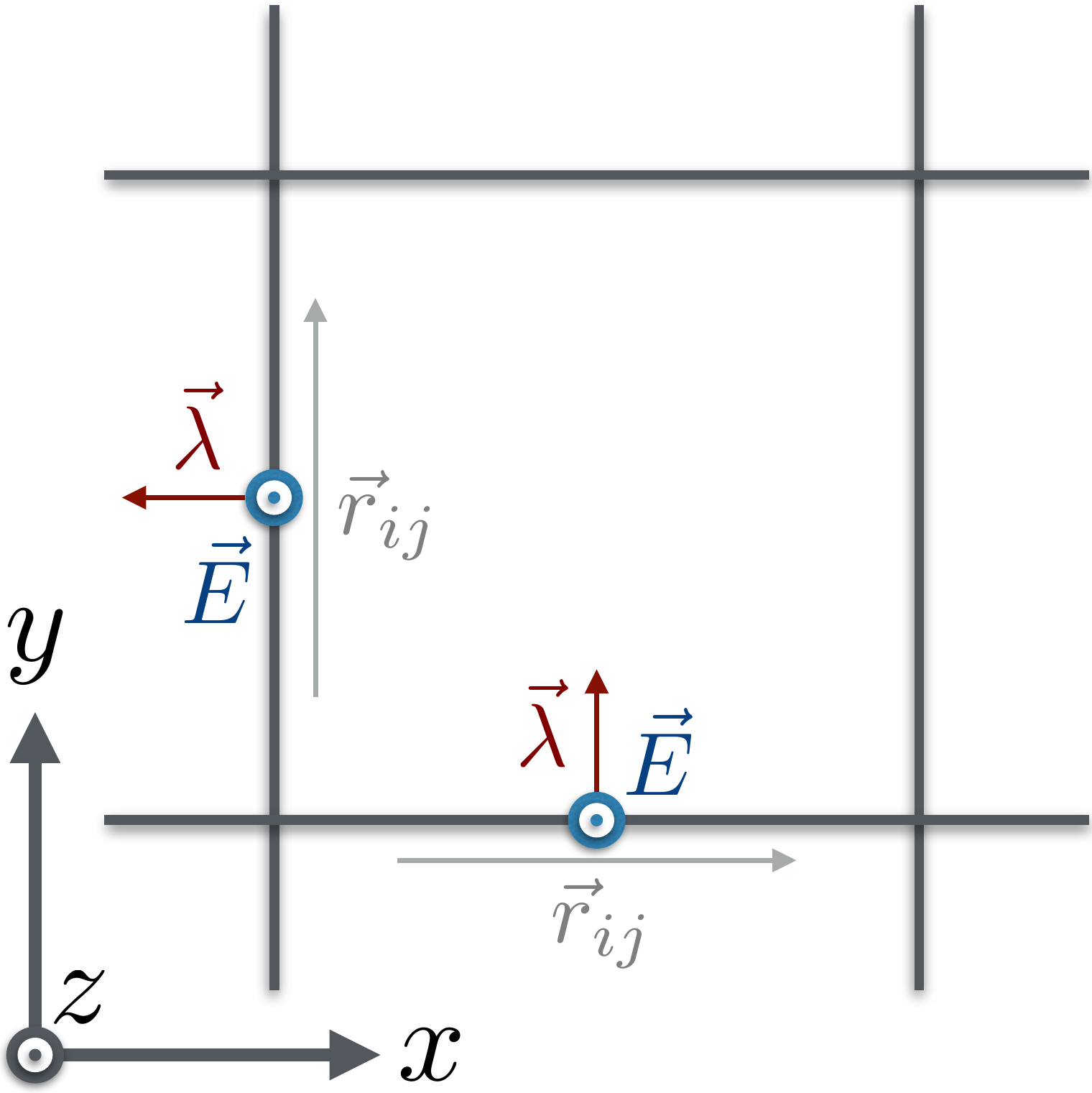}
\caption{Rashba spin orbit coupling arising from a perpendicular electric field. Hopping on each bond develops a spin-orbit term determined by the vector $\vec{\lambda}_{ij} = \vec{E}\times \vec{r}_{ij}$. This term is proportional to the amplitude of the electric field; we fix its strength to be $t_\lambda$, a tuning parameter. We only consider this term on nearest neighbour bonds.}
\label{fig.RSOC}
\end{figure}

We add Rashba spin-orbit coupling (RSOC) by means of spin-dependent hoppings\cite{Tang2014},
\begin{eqnarray}
\nonumber H_{RSOC} = - t_\lambda \sum_{\langle ij \rangle,\sigma} c_{i,\sigma}^\dagger 
\left\{\vec{\mathbf{\sigma}} \cdot 
\vec{\lambda}_{ij} \right\}_{\sigma\sigma'}
c_{j,\sigma'},
\label{eq.Hubbard_RSOC}
\end{eqnarray}
where $\vec{\mathbf{\sigma}}$ is the vector of Pauli matrices and $\vec{\lambda}_{ij}=\lambda ~\hat{z} \times \mathbf{r}_{ij} $ is the spin-orbit vector on each bond as shown in Fig.~\ref{fig.RSOC}. 
In the presence of this term, spin is no longer a good quantum number. For simplicity, we only include RSOC terms in the nearest neighbour hopping. 

We emphasize that the RSOC term here does not correspond to a microscopic realization of the SOC term described in Eq.~\ref{eq.LRSOC}. In the analysis on the sphere presented in Sec.~\ref{ssec.SOC}, the sign of the coefficient $\lambda$ is crucial to achieve a skyrmionic ground state. Using the wrong sign pushes the skyrmionic state to higher energies. In contrast, in our simulations of the Hubbard model with RSOC, we find similar solutions for both signs of $t_\lambda$. Nevertheless, both terms share the same physical origin, as they arise from inversion symmetry breaking.

We study this model using Bogoliubov-deGennes (BdG) mean field simulations, an approach that has been extensively used to study the effect of SOC on superconductivity\cite{Liu2012,Seo2013,Iskin2013,Xu2014,Qin2016}. On an $L\times L$ lattice, we obtain a $4L^2\times 4L^2$ matrix as the BdG Hamiltonian. 
All results presented here were obtained on a $24\times 24$ square lattice with periodic boundary conditions. 
We decouple the interaction term in both pairing and density channels. At each site, the superconducting and CDW order parameters, $\Delta_i = \langle c_{i,\uparrow} c_{i,\downarrow} \rangle$ and $\rho_i = (-1)^i \langle \hat{n}_{i,\uparrow} + \hat{n}_{i,\downarrow} - 1 \rangle$, are determined self-consistently. The quantity $\rho_i$ encodes checkerboard CDW order, given by the staggered deviation from half-filling. The explicit form of the gap and number equations are given in Ref.~\onlinecite{Iskin2013}. Vis-\`a-vis previous studies of s-wave pairing with SOC, our problem here has an additional ingredient in the form of an orbital magnetic field. This is introduced via Peierls' factors in the hopping amplitudes, $t$, $t'$ and $t_\lambda$, following Ref.~\onlinecite{KarmakarPRB2017}. For simplicity, we assume a uniform magnetic field, assuming a strongly type-II superconductor with $\kappa \gg 1$. This approach allows for studying vortex lattices, with vortex cores showing competing CDW order\cite{KarmakarPRB2017}.

\begin{figure}
\includegraphics[width=\columnwidth]{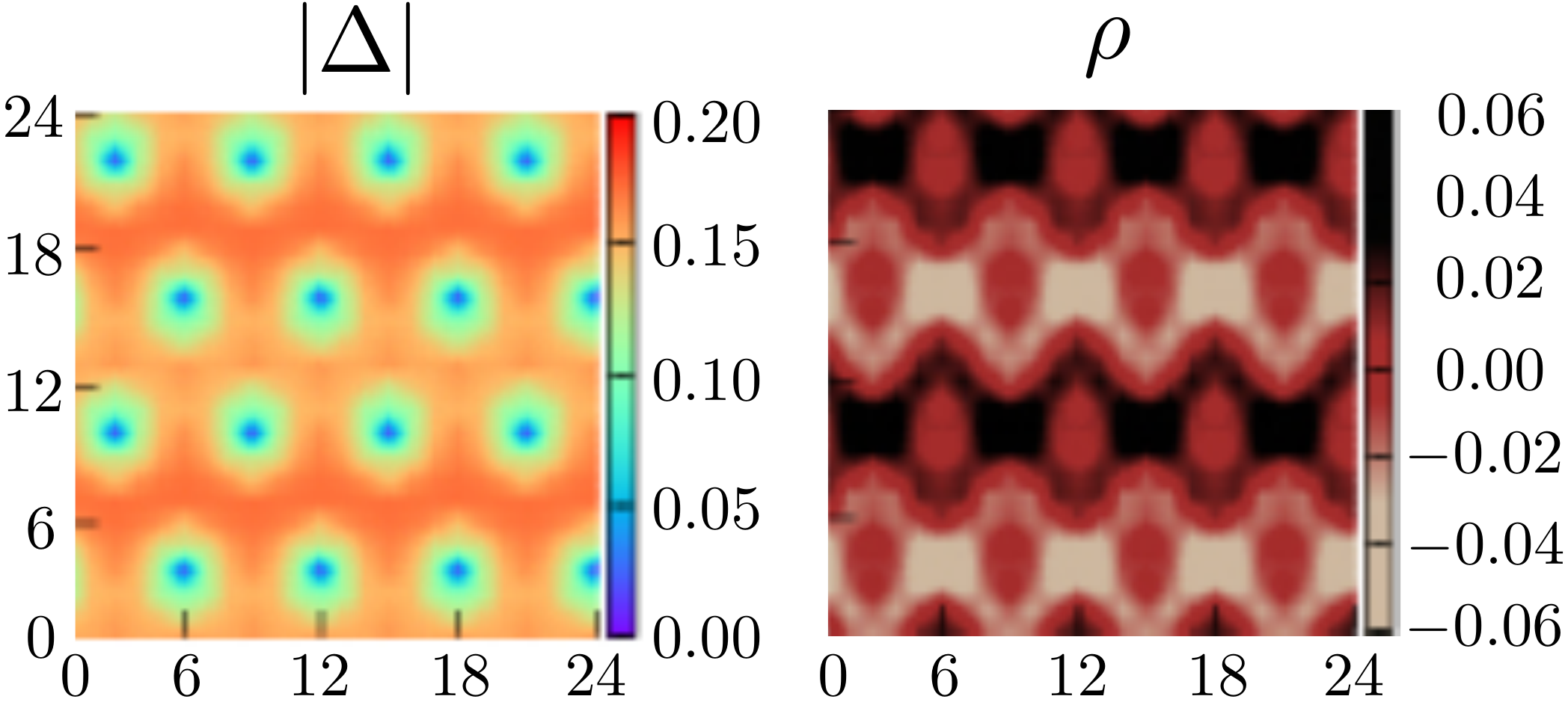}
\caption{Skyrmion-like stripe in the Hubbard model for $U=2t$, $t_\lambda=0.5$ and a field corresponding to sixteen vortices. Superconducting amplitude as a function of position (left) shows clear vortices. The CDW order parameter (right) has maximum amplitude at vortex cores, but alternates in sign to form stripes.}
\label{fig.U2_stripe}
\end{figure}

\begin{figure}
\includegraphics[width=\columnwidth]{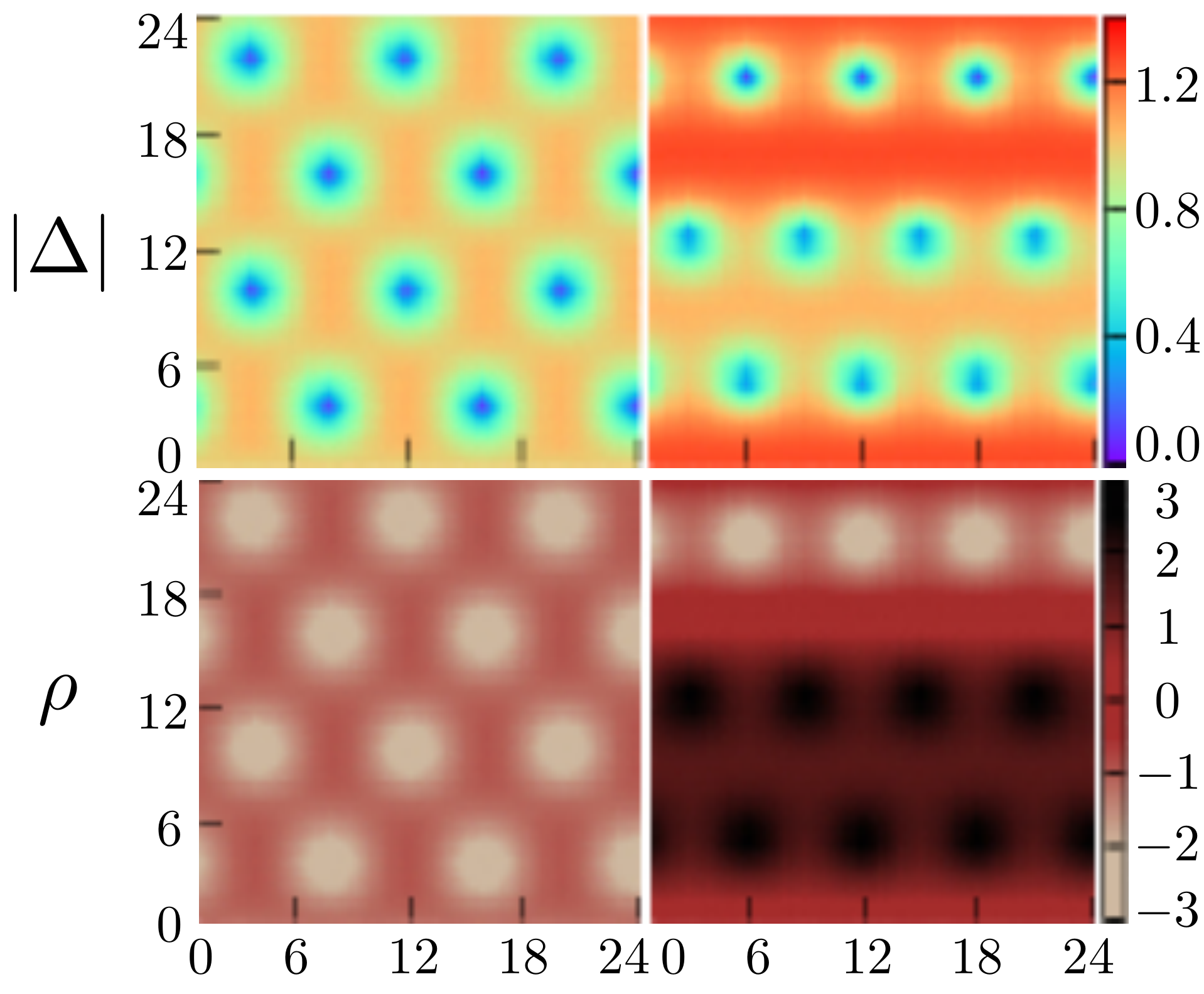}
\caption{Two solutions obtained for $U=4$, $t_\lambda =0.5$ at a flux corresponding to 12 vortices. 
A regular supersolid solution is shown at left, with panels showing superconducting amplitude (top left) and the CDW order parameter (bottom left). A domain-wall solution is shown in the corresponding panels on the right. We have coherent superconducting order, but CDW order is disrupted at a domain wall. These two solutions have the same free energy within our simulation resolution. }
\label{fig.U4_domainwall}
\end{figure}

\begin{figure}
\includegraphics[width=\columnwidth]{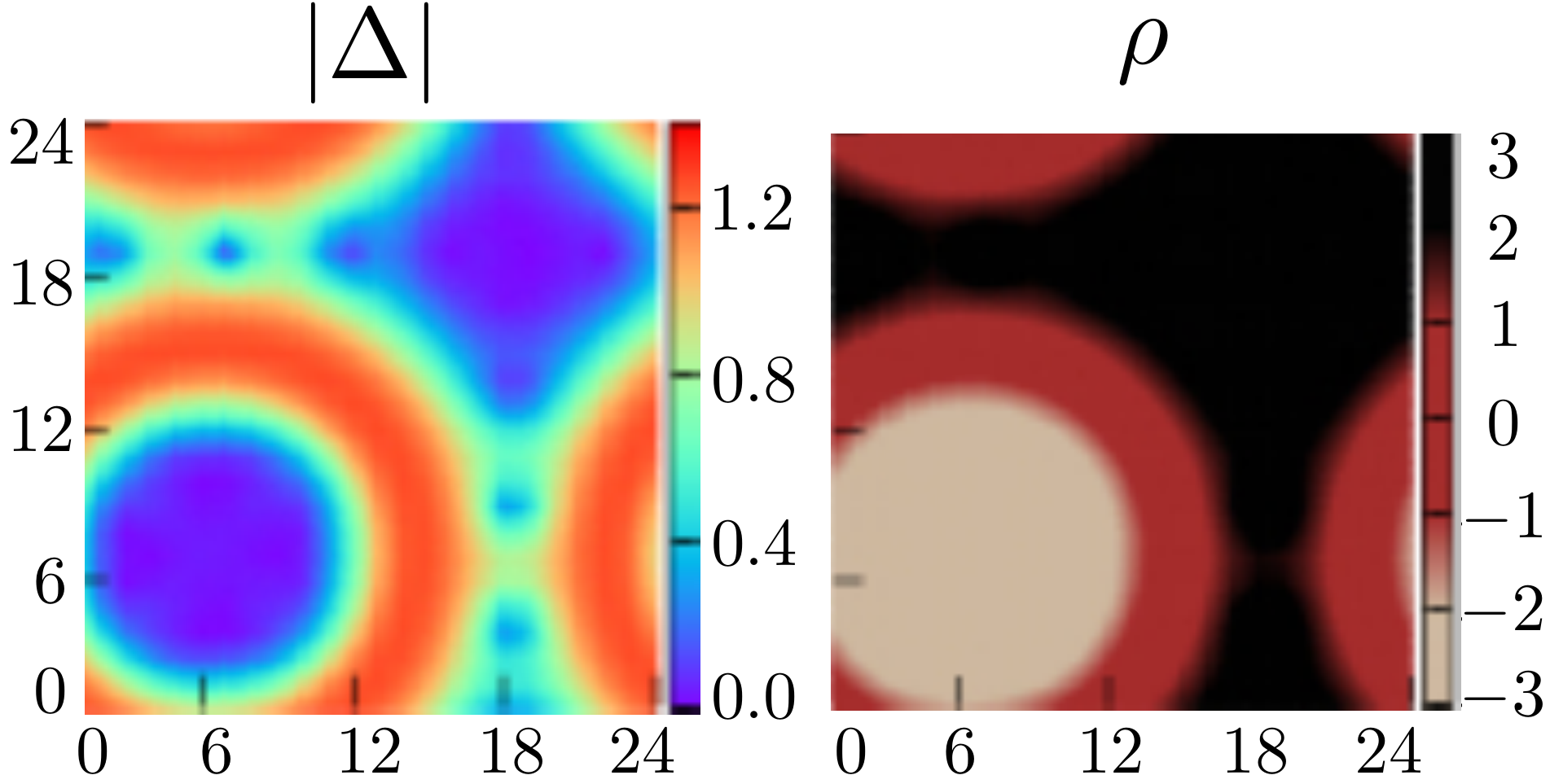}
\caption{A circular domain wall solution. The parameters used are $U=4$, $t_\lambda=0.5$ with a large flux corresponding to 20 vortices.}
\label{fig.U4_circular_domainwall}
\end{figure}

\begin{figure*}
\includegraphics[width=1.5\columnwidth]{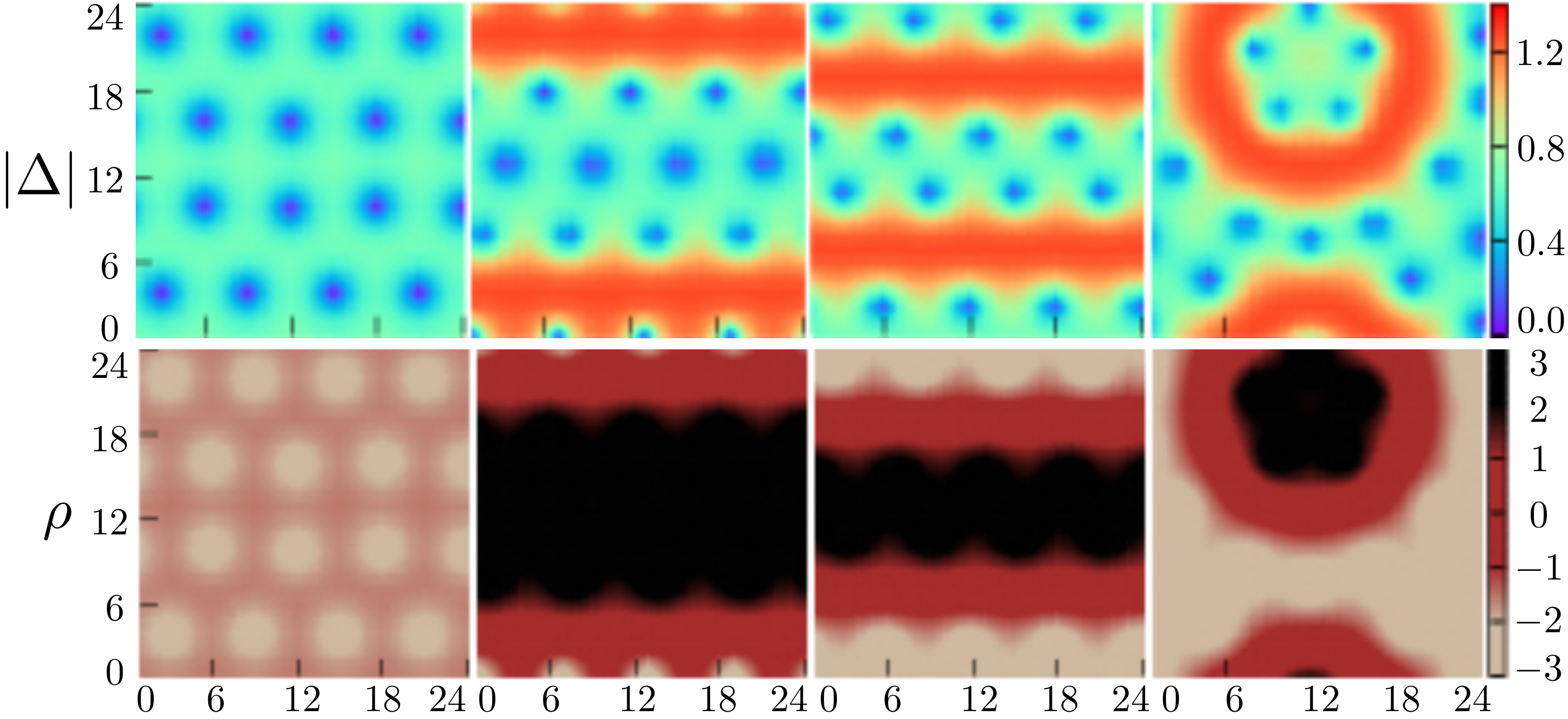}
\caption{Mean field solutions for $U=4$, $t_\lambda=0.5$ at a flux corresponding to 16 vortices. A regular supersolid solution is shown at left. The central panels show two solutions, both containing domain walls in CDW order. The panels on the right show a circular domain wall. All four solutions have comparable energies, with the differences being smaller than our resolution. }
\label{fig.U4_multiplesolutionsl}
\end{figure*}

\subsection{Solutions with skyrmion-like character}
In the absence of spin orbit coupling, sufficiently high magnetic fields give rise to `supersolid' solutions with coexisting superconductivity and checkerboard CDW order\cite{KarmakarPRB2017}. At very strong coupling, RSOC does not alter the mean field results. However, for intermediate coupling strengths ($U \lesssim 4t$), we find new solutions with interesting order parameter textures. In this regime, the Hubbard model does not realize the $SO(3)$ theory as the uniform spin length constraint is not always satisfied. A soft-spin model may describe the physics here\cite{Alama1999}. Nevertheless, we find solutions that bear similarities to the case of the sphere discussed in Sec.~\ref{sec.sphere} above.

A one-dimensional version of the skyrmion-like solution on the sphere is shown in Fig.~\ref{fig.U2_stripe}. This solution is obtained at $U=2t$ and $t_\lambda = 0.5t$, with a field corresponding to twelve vortices. Without RSOC, this parameter set gives a regular supersolid. However, with RSOC, we do not find a supersolid solution - all initial conditions in our simulations lead to solutions such as that shown in Fig.~\ref{fig.U2_stripe}. Here, each vortex core develops CDW order. However, the vortices arrange themselves into stripes with CDW character alternating from one stripe to the next. 
Along the stripe direction, vortices gain from energy lowering due to stronger CDW ordering. In the transverse direction, we have strong gradients with oscillations in the CDW character. Evidently, this is favoured by the RSOC term -- in a fashion that is similar to the case of the $SO(3)$ theory on the sphere discussed in Sec.~\ref{sec.sphere}. 
In the language of pseudospins, vortices along a stripe can be thought of as merons while vortices on the neighbouring stripe constitute anti-merons. We find this meron-antimeron stripe order to be a robust phenomenon over a range of parameter values.

\subsection{Loss of CDW rigidity}

In a large parameter regime, we find that RSOC destabilizes CDW order. In Fig.~\ref{fig.U4_domainwall}, we compare two mean field solutions found for $U=4t$ and $t_\lambda=0.5t$, with a magnetic field corresponding to twelve vortices. The panels on the left show a regular supersolid solution where both superconductivity and CDW order are strong. This solution approximately satisfies the uniform spin length constraint of the $SO(3)$ theory. 
This is to be contrasted with the panels on the right. Here, superconductivity remains strong. However, CDW order shows a clear domain wall, separating regions of opposite CDW order. The spin length constraint is violated on the domain wall -- this is a common feature in all our domain-wall solutions at intermediate coupling. Surprisingly, these two solutions are very close in free energy, with the difference being smaller than the energy resolution in our simulations ($\delta E \approx U/L^2$). This suggests that domain walls have low energy cost in the presence of RSOC. This indicates loss of CDW rigidity, as any small perturbation can induce domain walls.

We also find domain walls in circular geometries as shown in Fig.~\ref{fig.U4_circular_domainwall} for $U=4t$, $t_\lambda=0.5t$ and a field corresponding to twenty vortices. A circular region with predominant CDW character is surrounded by a superconducting ring. The ring can be thought of as a higher order vortex as it supports a large phase gradient. The enclosed CDW region forms an ordered vortex core. Outside the ring, we find CDW order of opposite character. This bears a strong similarity to the antiparallel solution on the sphere in Fig.~\ref{fig.antiparallel}, with two regions of opposite CDW order separated by a superconducting boundary. In pseudospin language, this resembles a higher order skyrmion.
We also find a regular supersolid solution for this parameter set, with lower free energy than the circular domain wall solution. However, the free energy difference is very small, $\approx 0.04t$ per site. 

Our results put together indicate loss of CDW rigidity. This is illustrated in Fig.~\ref{fig.U4_multiplesolutionsl} which shows multiple mean field solutions obtained for $U=4t$, $t_\lambda=0.5t$ and a field corresponding to sixteen vortices. The figure shows a regular supersolid, two domain wall solutions and a circular domain wall solution. Surprisingly, all these states are comparable in free energy, with the differences being less than our energy resolution. This indicates that there is essentially no cost for domain wall formation in CDW order. Small perturbations arising from thermal fluctuations or defects can easily proliferate domain walls. As a consequence, we argue that CDW correlations cannot develop over large length scales in spite of strong local ordering. In contrast, superconductivity survives with long ranged coherence. In such a system, probes such as X-ray diffraction will see diffuse peaks corresponding to short-ranged CDW order.

\section{Discussion}
We have presented a study of SOC-induced textures in superconductors with competing orders. We first demonstrate a skyrmion-like configuration on a sphere, within the $SO(3)$ theory of competing orders. This is a topologically stable solution consisting of two vortices with anti-correlated core order.
A term associated with inversion-symmetry-breaking stabilizes this configuration, indicating that SOC may favour skyrmionic textures. Moving to the two-dimensional plane, we study the attractive Hubbard model with RSOC hoppings. At intermediate coupling, for some parameters, we find skyrmion-like stripes with alternating CDW character. Over a wide parameter range, we find multiple mean field solutions which are comparable in free energies. The solutions correspond to spatially modulated CDW order due to RSOC-induced softening of domain walls. We argue that CDW loses its rigidity as a consequence.

In the underdoped cuprates, competition between charge order and superconductivity has been explained in terms of non-linear sigma models with an extended order parameter vector\cite{Meier2013,Nie2015,Caplan2015,Wachtel2014,Montiel2017}. For example, an $SO(6)$ theory has been proposed combining superconductivity and two incommensurate CDW's\cite{Hayward2014}. As the $SO(3)$ theory has the same field theory structure, the results presented may shed light upon the physics seen in the cuprates. 
In particular, our results on the Hubbard model with RSOC could explain why no cuprate shows long-ranged charge order. With low energy cost for domain wall formation, even weak disorder can lead to an irregular CDW pattern with several domain walls.
The cuprates may have reasonably strong SOC, as they typically contain heavy elements such as La, Bi and Yb. In addition, YBCO\cite{Briffa2016} (the cuprate with the strongest charge order) and BSSCO-2212 lack inversion symmetry about the Cu-O plane. This allows for RSOC terms within each plane\cite{Gotlieb2018,Raines2018} which can destroy CDW rigidity. In such a scenario, we will have robust local CDW correlations but no long range order. This could explain several experimental observations on the underdoped cuprates.

The physics of skyrmions has been extensively studied in magnetism, in both ferromagnets and antiferromagnets. Our study opens the door to skyrmion-like defects in superconductors. Experiments on the cuprates may be able to uncover skyrmionic textures around vortices. Probing their precise structure, response to perturbations, dynamics, etc., can lead to interesting and potentially useful insights. In particular, textured halos of competing order may provide a new handle to manipulate vortices.

\appendix
\section{Euler-Lagrange equation on the sphere}
\label{app.ELeqn}
We derive the saddle point equations for the action given in Eq.~\ref{eq.LG}, defined on a sphere. We first consider the northern hemisphere ($\theta <\pi/2$). The vector potential and the order parameter profile in this region are given in Eqs.~\ref{eq.vecA} and \ref{eq.Deltadef} respectively. The CDW order parameter, as defined in Eq.~\ref{eq.rho}, can take two possible signs. Regardless of this choice, the action takes the form
\begin{eqnarray} 
 \mathcal{L}\Big(f(\theta),f'(\theta),\theta\Big) &=&\left| \Big( \nabla - i \frac{(1 - \cos\theta)}{r \sin\theta}\hat{\phi} \Big) f(\theta) e^{i \phi}\right|^{2} \nonumber \\
\nonumber &+& \left| \nabla\sqrt{c^2 - f(\theta)^{2}}\right|^{2}- \frac{2}{\chi} f(\theta)^{2} \\
 &-& \frac{2}{\chi}(1 - \delta)(c^2 - f(\theta)^{2}). 
\end{eqnarray} 
We note that there is no explicit dependence on $\theta$. We define $\tilde{f} = f/c$ and scale the action by $R^2/c^2$. After several simplifications, the action (ignoring a constant shift) takes the form
\begin{eqnarray}\label{Eq. 8} \mathcal{L} \Big(\tilde{f},\tilde{f}' \Big) \!=
\! \frac{1}{(1 - \tilde{f}^{2})}
\left(\frac{\partial \tilde{f}}{\partial \theta}\right)^{2} 
\!+\!  \Big(
\cot^{2}\theta 
-\frac{1}{\xi^{2}} \Big) \tilde{f}^{2} + \frac{1}{\xi^{2}}.~~~
\label{eq.action}
\end{eqnarray}
This action is extremized by the Euler-Lagrange equation, $\frac{d}{d\theta}\left( \frac{\partial \mathcal{L}}{\partial 
\tilde{f}'}  \right)= \frac{\partial \mathcal{L}}{\partial 
\tilde{f}}$. This leads to the saddle point equation given in Eq.~\ref{eq.EL}. Once a saddle point solution is found, its free energy can be found by integrating the free energy density (the action) in Eq.~\ref{eq.action} over the sphere.

In the southern hemisphere, the vector potential and the order parameter take different forms. Nevertheless, the action turns out to be given by the same expression as in Eq.~\ref{Eq. 8}. As a result, the Euler-Lagrange equation also takes the same form as in the northern hemisphere. As a result of this simplification, we can solve Eq.~\ref{eq.EL} for the saddle point solution over the entire sphere.

\section{Setting up the shooting method}
\label{app.shooting}

As described in the main text, we use the shooting method to obtain saddle point solutions. This approach converts a boundary value problem into an initial value problem, where we specify $f(\theta)$ and its derivative at a boundary ($\theta=0$). To justify this, we consider $\theta \rightarrow 0$, where the saddle point equation reduces to
\begin{equation}
f'(\theta) -\frac{1}{\theta}f(\theta) = 0.
\end{equation}
This has linear solutions, $f(\theta) = \alpha~ \theta$. Guessing $f'(\theta)$ at $\theta=0$ corresponds to guessing the coefficient $\alpha$.

Strictly speaking, we must solve for $f(\theta)$ independently in the two hemispheres. On symmetry grounds, the solution must be continuous at the equator, i.e., we must have $f(\theta \rightarrow \pi/2^+) = f(\theta \rightarrow \pi/2^-)$. All derivatives must also be continuous at the equator, e.g. with $f'(\theta \rightarrow \pi/2^+) = f'(\theta \rightarrow \pi/2^-)$.
These conditions follow from reflection symmetry about the equator. 
They allow us to consider $f(\theta)$ as one continuous function over the domain $\theta \in [0,\pi]$. On this basis, we use the shooting method over the entire sphere: we fix the function and its derivative at $\theta=0$ and demand that it satisfy $f(\pi)=0$.

\section{Role of SOC on the sphere}
\label{app.SOCsphere}
We describe the role of the SOC term in the SO(3) theory on the sphere. In Eq.~\ref{eq.Deltadef}, we have used the symmetry of the sphere to constrain the form of the superconducting order parameter. We first consider the northern hemisphere. 
The perturbation $L_{SOC}$, given in Eq.~\ref{eq.LRSOC}, reduces to 
\begin{eqnarray}
\nonumber L_{SOC} &=&
 i \lambda ~ \hat{r}\cdot \Big[
 \mathbf{\nabla}  \Delta \times
 \mathbf{\nabla}  \Delta^*  \\
 &-& 
 \frac{2ie}{\hbar c}  \left\{
\Delta( \mathbf{A}  \times  \mathbf{\nabla} \Delta^*) +  \Delta^* (\mathbf{\nabla} \Delta \times  \mathbf{A} ) \right\}
\Big].~~
\end{eqnarray} 
Using the form of $\Delta(\theta,\phi)$ and $\vec{A}$ in the northern hemisphere, we find that the terms involving the vector potential cancel. Conveniently for us, this cancellation occurs in the southern hemisphere as well. We are left with 
\begin{equation}
L_{SOC} =  \left\{   
\begin{array}{c}
\frac{2\lambda}{\sin(\theta)} ~f(\theta)f'(\theta), ~\theta\leq \pi/2 \\
\frac{-2\lambda}{\sin(\theta)} ~ f(\theta)f'(\theta),~ \theta> \pi/2 \\
\end{array}
\right..
\label{eq.SOC_sphereb}
\end{equation}
Remarkably, this additional term does not change the Euler-Lagrange equation. Its contribution to the free energy is 
\begin{eqnarray}
\int \sin\theta d\theta d\phi ~ \Big(\frac{\pm2\lambda}{\sin\theta} \Big) f(\theta)f'(\theta)
= 
\pm 2\pi \lambda \int d\theta \frac{d\{ f^2(\theta)\}}{d\theta}, \nonumber
\end{eqnarray}
a total derivative! Naively, we may disregard this term as the order parameter vanishes at the boundaries, $\theta=0$ and $\theta=\pi$. However, we should consider the problem separately in the two hemispheres with each contributing a boundary term at $\theta=\pi/2$. As the free energy contribution has opposite signs in the two hemispheres, these two boundary term add up. This leads to the free energy correction given in Eq.~\ref{eq.FE_SOC}.

As discussed in the main text, this term provides different energy corrections to the parallel and antiparallel solutions. The physical origin of this distinction can be deduced from Eq.~\ref{eq.SOC_sphereb}. The contribution to the free energy density is proportional to $f'(\theta)$. As the antiparallel solution has a stronger $\theta$-derivative (compare Figs.~\ref{fig.antiparallel} and \ref{fig.parallel}), it is favoured by the SOC term.


\bibliographystyle{apsrev4-1}
\bibliography{HubbardSO3}

\end{document}